\documentclass[cits]{PoS}

\title{Z(2)-symmetric center vortex model with a first-order
deconfinement transition}

\ShortTitle{Z(2)-symmetric center vortex model with a first-order
deconfinement transition}

\author{\speaker{M.~Engelhardt} and B.~Sperisen\\
        Physics Department, New Mexico State University\\
        Las Cruces, NM 88003, USA\\
        E-mail: \email{engel@nmsu.edu}}

\abstract{A random vortex world-surface model for the infrared sector of
$Sp(2)$ Yang-Mills theory is constructed. The $Sp(2)$ gauge group, while
allowing for the same set of center vortex fluxes as the $SU(2)$ gauge
group, induces a significantly different dynamics on those vortex fluxes,
which manifests itself in a first-order deconfinement phase transition.
As shown by the construction presented here, a new vortex effective
action term which can be interpreted in terms of a vortex stickiness can
be used to drive the deconfinement transition towards first-order
behavior. The available data from lattice $Sp(2)$ Yang-Mills theory
are reproduced quantitatively and, in addition, predictions for the
behavior of the spatial string tension at high temperatures are
presented.}

\FullConference{XXIVth International Symposium on Lattice Field Theory\\
                July 23-28, 2006\\
                Tucson, Arizona, USA}

\begin{document}

\section{Introduction}
Recently, efforts have been undertaken to gain further insight into
confinement mechanisms by studying Yang-Mills theories based on gauge
groups other than $SU(N)$. In particular, the $G(2)$, $Sp(2)$ and $Sp(3)$
groups have been studied in detail \cite{g2,spn,g21o,pepelat05}. One
interesting comparison made possible by that work is the one between the
$Sp(2)$ and the $SU(2)$ cases: Both of these groups have the same center
and also induce the same set of center vortex degrees of freedom in the
respective Yang-Mills theories\footnote{Note that the center vortex
fluxes allowed for by a gauge group $G$ are determined by the first
homotopy group of the gauge group after factoring out its center $Z_G $,
i.e., $\Pi_{1} (G/Z_G )$. In more physical terms, specifically for $SU(2)$
and all $Sp(N)$ groups, the chromomagnetic flux carried by center vortices
is quantized such as to contribute a center phase $(-1)$ to any Wilson
loop to which it is linked.}. Nevertheless, $SU(2)$ Yang-Mills theory
displays a second-order deconfinement phase transition, whereas the
$Sp(2)$ case displays a first-order transition. This observation,
among others, led to the conclusion in \cite{spn,g21o,pepelat05} that the
center plays no role in determining the order of the deconfinement phase
transition, which is instead decisively influenced by the size of the
gauge group.

Formulated slightly more precisely, the conclusion which can be drawn
from the aforementioned work is that knowledge of the center of a gauge
group alone does not permit a prediction of the order of the deconfinement
transition. However, it should not be misconstrued as implying that the
center degrees of freedom of the theory are irrelevant for the physics
of confinement and the deconfinement transition. Indeed, knowledge of
the set of center vortex degrees of freedom contained in a Yang-Mills
theory does not suffice to specify its infrared effective vortex
description and the associated infrared phenomenology; one of course
must also give the effective action governing those vortex degrees of
freedom. Only then is the infrared description complete. Vortex models
based on the same set of degrees of freedom, but
with different effective actions, can lead to vastly different physics.
Accordingly, in the present investigation, an effective vortex model for
the infrared sector of $Sp(2)$ Yang-Mills theory is constructed. While
based on the same set of vortex degrees of freedom as the $SU(2)$ model
investigated in \cite{m1,m2,m3}, it displays a first-order deconfinement
phase transition, contrary to the latter case. Thus, the vortex picture
is perfectly consistent with the $SU(2)$-$Sp(2)$ comparison highlighted
above; in the vortex language, integrating out the (differently
sized) cosets in $SU(2)$ and $Sp(2)$ Yang-Mills theory yields
different effective actions for the respective vortex degrees of
freedom. These different effective actions lead, in particular, to
different behavior at the deconfinement transition. From this point of view,
thus, the physics of confinement and, in particular, the deconfinement
transition is still determined by infrared effective center vortex
dynamics; the distinction between the $SU(2)$ and $Sp(2)$ vortex models
arises at the level of the specific effective actions inherited from the
full gauge groups.

\section{Lattice Yang-Mills data}
The study \cite{spn} provides two relevant confinement characteristics
of $Sp(2)$ Yang-Mills theory, namely, the ratio of the deconfinement
temperature to the square root of the zero-temperature string tension,
$T_c /\sqrt{\sigma } $, and the latent heat at the transition $L_H $.
The latter corresponds to the discontinuity in the four-dimensional
action density\footnote{Since the symbol $s$ will be put to a different
use further below, the action density is denoted by $\bar{s} $ in the
following.} $\bar{s} $ at the first-order deconfinement transition, and is
given in \cite{spn} in lattice units, i.e., $L_H = a^4 \Delta \bar{s} $,
where $a$ denotes the lattice spacing. It should be noted that \cite{spn}
only gives $L_H $ quantitatively at one rather strong coupling,
$8/g^2 = 6.4643$; on the other hand, $T_c /\sqrt{\sigma } $ is available
for a number of couplings, including the extrapolation to the continuum
limit. The most consistent way to model the available data is to aim at
reproducing $Sp(2)$ Yang-Mills theory specifically at the coupling
$8/g^2 = 6.4643$, at which both $L_H $ and $T_c /\sqrt{\sigma } $ are
known (rather than using input from two different couplings, i.e.,
$8/g^2 = 6.4643$ in the one case and the continuum limit in the other).

For $8/g^2 = 6.4643$, the ratio of the deconfinement temperature to the
square root of the zero-temperature string tension is \cite{spn}
\begin{equation}
T_c /\sqrt{\sigma } = 0.59 \ .
\label{input1}
\end{equation}
On the other hand, identifying $L_H = a^4 \Delta \bar{s} $, one has
\cite{spn}
\begin{equation}
N_t (a^4 \Delta \bar{s} )^2 /4 =0.15
\end{equation}
for the action density discontinuity $\Delta \bar{s} $, where $N_t $ denotes
the extent of the lattice in the (Euclidean) time direction. Combining this
with the fact that, at $8/g^2 = 6.4643$, the deconfinement transition
occurs at $N_t =2$, i.e., the deconfinement temperature satisfies
$aT_c =1/2$, the lattice spacing can be eliminated, resulting in
\begin{equation}
\Delta \bar{s} / T_c^4 = 8.76 \ .
\label{input2}
\end{equation}
Relations (\ref{input1}) and (\ref{input2}) are used as input
data for the random vortex world-surface model to be constructed in
the following.

\section{Random vortex world-surface model}
Random vortex world-surface models describe the infrared, strongly
interacting regime of Yang-Mills theories on the basis of effective
gluonic center vortex degrees of freedom. A description in terms of such
degrees of freedom was initially suggested and studied in
\cite{hooft,aharonov,cornold,mack,olesen}; compelling motivation
for this picture is provided by more recent investigations
of the relevance of center vortices in the lattice Yang-Mills ensemble
\cite{jg1,jg2,tk1,df1,per,rb}, for a review, cf.~\cite{jg3}. Random vortex
world-surface models have been investigated both with respect to $SU(2)$ as
well as $SU(3)$ Yang-Mills theory \cite{m1,m2,m3,su3conf,su3bary,su3freee}.
They successfully reproduce the main features of the strongly interacting
vacuum. In the $SU(2)$ case, not only has a confining low-temperature phase
been obtained together with a second-order deconfinement phase transition as
temperature is raised \cite{m1}; also the topological susceptibility
\cite{m2,cw2,contvort,bruck} and the (quenched) chiral condensate \cite{m3}
of $SU(2)$ Yang-Mills theory are reproduced quantitatively. In the $SU(3)$
case, the deconfinement transition becomes weakly first order \cite{su3conf}
and a Y-law for the baryonic static potential results in the confining phase
\cite{su3bary}.

Recent efforts have concentrated on exploring the range
of applicability of random vortex world-surface models as the Yang-Mills
gauge group is varied. An investigation of the confinement properties
in a vortex model for $SU(4)$ Yang-Mills theory \cite{su4} revealed
signatures of Abelian magnetic monopoles, which are an intrinsic feature
of generic center vortex configurations due to vortex world-surface
non-orientability, beginning to influence the vortex dynamics for $SU(4)$
color. Contrary to the $SU(2)$ and $SU(3)$ cases, the vortex action cannot
be expressed purely in terms of world-surface properties anymore, but
acquires contributions attributable to the monopoles. Such a shift in the
dynamical characteristics is expected as the number of colors rises
\cite{jeffstef}; the model construction presented in \cite{su4} explicitly
confirms that expectation. The thrust of the present investigation is
similar in nature, as already discussed further above: To be able to
model the infrared sector of $Sp(2)$ Yang-Mills theory, new features need
to be introduced into the effective vortex action.

A practical computational framework for random vortex world-surface models
is achieved by composing vortex world-surfaces of elementary squares on a
hypercubic space-time lattice\footnote{The lattice spacing in
this approach is a finite physical quantity implementing the notion that
vortices possess a finite transverse thickness and must be a minimal
distance apart to be distinguished from one another.}. An ensemble of
random world-surfaces is generated by Metropolis Monte Carlo update,
where the closed nature of the surfaces (implied by the Bianchi constraint
on the chromomagnetic fields) is preserved by simultaneously updating
all six sides of an elementary three-dimensional cube at each step.
The weighting of world-surface configurations is determined by an action
of the symbolic form
\vspace{0.1cm}
\begin{equation}
S\ \ = \ \ c \ \times
\hspace{1.9cm} + \ \ \ \ s \ \times
\hspace{2cm} .
\label{voract}
\end{equation}
\vspace{-2.1cm}

\begin{figure}[h]
\centerline{\hspace{1.4cm} \includegraphics[width=1.7cm]{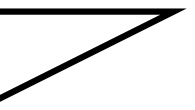}
\hspace{1.3cm} \includegraphics[width=1.53cm]{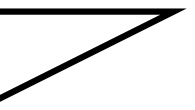} }
\end{figure}
\vspace{-0.3cm}

The first of these two action terms is a curvature term: Each instance
of two vortex elementary squares sharing a link without lying in the
same plane costs an action increment $c$, thus penalizing vortices
``going around a corner''. For the $SU(2)$ and $SU(3)$ cases
studied in \cite{m1,m2,m3,su3conf,su3bary,su3freee}, this single term
was sufficient to achieve a phenomenologically viable model for the
infrared sector of the corresponding Yang-Mills theory.

By contrast, to arrive at a model for the infrared sector of $Sp(2)$
Yang-Mills theory, additional dynamics must be introduced; otherwise,
the finite-temperature deconfinement phase transition remains second
order. The second term in (\ref{voract}) can be interpreted as follows:
Each instance of more than two vortex elementary squares being attached
to a link is weighted\footnote{This includes both the case of four as
well as the case of six elementary squares being attached to the link.
In general, one can associate these two cases with different action
increments; however, throughout the present investigation, they were weighted,
ad hoc, with the same action increment.} by an action increment $s$. Thus,
for negative coefficient $s$, vortex world-surfaces are encouraged to
intersect along whole lines in space-time. This means that, when two vortex
flux lines meet, they tend to maintain their contact over longer distances.
In this sense, the second action term in (\ref{voract}) can be viewed as
implementing a ``stickiness'' of the vortices. This choice of action
is motivated by the experience with the $SU(4)$ random vortex world-surface
model \cite{su4}. Also in that case, it is necessary to enhance the
first-order character of the deconfinement phase transition. This is
achieved by facilitating vortex branching, a process possible in the
$SU(4)$ case due to the existence of two types of center flux. While
the $Sp(2)$ case discussed here does not allow for branching since
there is only one type of center flux, the stickiness term in (\ref{voract})
is similar to a branching term in that it encourages more than two vortex
elementary squares being attached to a link (for negative $s$). For this
reason, such an action term seems a promising choice, which indeed is
vindicated by the numerical results described below.

\section{Locating the physical point}
Having generated a random vortex world-surface ensemble according to the
model defined above, one can measure the action density and the string
tension (using the fact that Wilson loops acquire a center phase $(-1)$
for every instance of a vortex piercing an area spanned by the loop).
Since the lattice spacing is a finite physical quantity, only
discrete temperatures can be realized at any given set of coupling
parameters $c$ and $s$ in (\ref{voract}). In general, therefore, the
deconfinement phase transition is not directly accessible at the
physical set of coupling parameters. Instead, one studies several unphysical
sets of $c$, $s$ which do realize the transition (as evidenced by a
double peak in the action density distribution) on lattices extending a
varying number $N_t $ of spacings in the (Euclidean) time direction.
The gathered data can then be interpolated to obtain, e.g., the latent
heat at the physical point in the space of coupling constants. The
simplest such interpolation scheme is obtained by using $N_t =1,2$,
equivalent to $aT_c =1$ and $aT_c =0.5$; Table \ref{physpt} displays
corresponding data taken at suitable coupling parameters $c$, $s$.

\begin{table}[h]
\begin{center}
\begin{tabular}{|c||c|c|c|c|}
\hline
$aT_c $ & $c$ & $s$ & $\Delta \bar{s} / T_c^4 $ & $T_c /\sqrt{\sigma } $ \\
\hline \hline
1 & 0.3394 & -1.24 & 0.014 & 0.816 \\
\hline
0.5 & 0.5469 & -1.99 & 13 & 0.474 \\
\hline
\end{tabular}
\end{center}
\caption{Sets of coupling parameters $c$, $s$ realizing the deconfinement
phase transition on lattices with $N_t =1,2$, together with measurements
of the latent heat and the ratio of the deconfinement temperature to the
square root of the zero-temperature string tension.}
\label{physpt}
\end{table}

Note that the requirement of realizing the deconfinement transition alone
of course does not fix both $c$ and $s$ simultaneously; the particular
sets of coupling parameters quoted in Table \ref{physpt} were singled
out by the additional requirement that linear interpolation of the data must
yield the physical point, at which both the relations (\ref{input1}) and
(\ref{input2}) are satisfied. Indeed, linearly interpolating
$\Delta \bar{s} / T_c^4 $ as a function of $T_c / \sqrt{\sigma } $, one
obtains $\Delta \bar{s} / T_c^4 =8.76$, as required by (\ref{input2}), for
$T_c / \sqrt{\sigma } \approx 0.59$, cf.~(\ref{input1}).
Likewise, linearly interpolating $c$ and $s$ as functions of
$T_c / \sqrt{\sigma } $ yields the physical set of coupling parameters
\begin{equation}
c = 0.479 \ \ \ \ \ \ \ \ \ \ \ \ \ \ \ s = -1.745 \ .
\label{physcs}
\end{equation}
Finally, one can also interpolate $aT_c $ as a function of
$T_c / \sqrt{\sigma } $, which yields
\begin{equation}
aT_c = 0.663
\label{phystc}
\end{equation}
at the physical point. Thus, contrary to the $SU(N)$ vortex models
\cite{m1,su3conf,su4}, the physical point in the present case is not
near one of the measured data sets. In all $SU(N)$ cases studied, the
physical point is very near the $N_t =2$ data set, and the uncertainty
engendered by the interpolation is consequently very small. Here, by
contrast, the uncertainty inherent in the interpolation is substantial
and it is useful to consider consistency checks. One simple such check
can be made as follows: While deconfinement transition data are not
directly accessible at the physical point, the zero-temperature string
tension $\sigma $ is. One can measure $\sigma a^2 $ directly for the
physical set of couplings (\ref{physcs}) and combine this with
(\ref{phystc}) to obtain an additional determination of
$T_c / \sqrt{\sigma } $. This indeed again yields
$T_c / \sqrt{\sigma } =0.59$, buttressing the interpolation procedure
used above. Another possibility, which will be reported in detail
elsewhere \cite{sp2full}, lies in using a larger data set, with
$N_t =1,2,3$, for the interpolation. This yields deviations of the order
of $10\, $\% in the determination of the physical point, giving another
indication of the amount of uncertainty. It should, however, be kept in
mind that using $N_t =1,2,3$ is not necessarily more accurate than just
using $N_t =1,2$ data; after all, $N_t =3$ is rather far removed from
the physical point and may thus easily introduce systematic distortions
into the results rather than rendering them more accurate.

Finally, having determined the physical set of coupling parameters
(\ref{physcs}), one can predict further nonperturbative characteristics
of $Sp(2)$ Yang-Mills theory, such as the behavior of the spatial string
tension $\sigma_{S} $ at high temperatures. Corresponding measurements
for different $N_t $ using (\ref{physcs}) are displayed in Table
\ref{spats}, where $N_t $ has been translated into $T/T_c $ using
(\ref{phystc}).

\begin{table}[h]
\begin{center}
\begin{tabular}{|c||c|c|c|}
\hline
$T/T_c $ & 0.50 & 0.75 & 1.51 \\
\hline \hline
$\sigma_{S} (T) / \sigma_{S} (T=0)$ & 1.00 & 1.02 & 1.36 \\
\hline
\end{tabular}
\end{center}
\caption{Predictions for the behavior of the spatial string tension
$\sigma_{S} $ at high temperatures.}
\label{spats}
\end{table}

These predictions can be used to test the validity of the present random
vortex world-surface model construction by comparing with corresponding
measurements in $Sp(2)$ lattice Yang-Mills theory.

\section{Conclusions}
By introducing a vortex ``stickiness'' into the action of an effective
$Z(2)$-symmetric random vortex world-surface model, it is possible to
drive its finite-temperature deconfinement phase transition towards
first-order behavior. This is necessary if such a model in particular is
to replicate the infrared phenomenology of $Sp(2)$ Yang-Mills theory.
Indeed, within the present investigation, it proved possible to reproduce
available data from lattice $Sp(2)$ Yang-Mills theory quantitatively
by appropriate choice of the curvature and stickiness coupling parameters.

This successful reconstruction of $Sp(2)$ Yang-Mills confinement
characteristics within the random vortex world-surface model underscores
that the $SU(2)$ -- $Sp(2)$ comparison discussed at the outset does not
contradict the vortex picture of Yang-Mills vacuum dynamics; while
the two gauge groups indeed engender the same vortex {\em topology}, they
lead to different vortex {\em effective actions} (induced by integrating out
their differently sized cosets), and thus naturally to different behavior at
the deconfinement phase transition.

In addition, the random vortex world-surface model constructed here yields
predictions for the behavior of the spatial string tension at high
temperatures, cf.~Table \ref{spats}. Comparison with corresponding
measurements within $Sp(2)$ lattice Yang-Mills theory could be used to
further test the validity of the model.

\acknowledgments
This work was supported by the U.S.~DOE under grants DE-FG03-95ER40965
(M.E.) and DE-FG02-94ER40847 (B.S.).

\end{document}